%% file: wsdm-2017-mise-en-scen.tex
\documentclass{sig-alternate-05-2015}
\usepackage{multirow}
\usepackage{makeidx}  
\usepackage{amsmath}
\usepackage{amssymb}
\usepackage{color}
\usepackage{graphicx}
\usepackage{xspace}
\usepackage{enumerate}
\newcommand{\norm}[1]{\left\lVert#1\right\rVert}
\newcommand{\mise}{mise-en-sc\`ene\xspace}
\newcommand{\Mise}{Mise-en-Sc\`ene\xspace}
\newcommand{\mpeg}{MPEG-7\xspace}
\newcommand{\deep}{deep-learning networks\xspace}
\newcommand{\Deep}{Deep-Learning\xspace}

\newcommand{\argmin}{\operatornamewithlimits{argmin}}

\usepackage{etoolbox}
\makeatletter
\patchcmd{\maketitle}{\@copyrightspace}{}{}{}
\makeatother

\usepackage{enumitem}
\setlist[enumerate,1]{%
  label=\arabic*.,
}
\newlist{inlinelist}{enumerate*}{1}
\setlist*[inlinelist,1]{%
  label=(\roman*),
}

\usepackage{csquotes}

\begin{document}







\title{Using \Mise Visual Features based on MPEG-7 and Deep Learning for Movie Recommendation}
%
%
%
%
%

\numberofauthors{4} 
\author{
\alignauthor
Yashar Deldjoo\\
       \affaddr{Politecnico di Milano}\\
       \affaddr{Via Ponzio 34/5}\\
       \affaddr{20133, Milan, Italy}\\
       \email{yashar.deldjoo@polimi.it}
\alignauthor
Massimo Quadrana\\
       \affaddr{Politecnico di Milano}\\
       \affaddr{Via Ponzio 34/5}\\
       \affaddr{20133, Milan, Italy}\\
       \email{massimo.quadrana@polimi.it}
\alignauthor
Mehdi Elahi\\
       \affaddr{Politecnico di Milano}\\
       \affaddr{Via Ponzio 34/5}\\
       \affaddr{20133, Milan, Italy}\\
       \email{mehdi.elahi@polimi.it}
\and  
\alignauthor Paolo Cremonesi\\
       \affaddr{Politecnico di Milano}\\
       \affaddr{Via Ponzio 34/5}\\
       \affaddr{20133, Milan, Italy}\\
       \email{paolo.cremonesi@polimi.it}}

\maketitle

\input{abstract}

\input{acm-generated-code}

%
%



\input{introduction}

\input{related}

\input{method}

\input{results}

\input{discussion}

\input{conclusion}

\section{Acknowledgments}
This work is supported by Telecom Italia S.p.A., Open Innovation Department, Joint Open Lab S-Cube, Milan.
The work has been also supported by the Amazon AWS Cloud Credits for Research program.

%
\bibliographystyle{abbrv}
\bibliography{wsdm2017}   
 
\end{document}

%% file: abstract.tex
\begin{abstract}

Item features play an important role in movie recommender systems, where recommendations can be generated by using explicit or implicit preferences of users on traditional features (attributes) such as tag, genre, and cast. 
Typically, movie features are human-generated, either editorially (e.g., genre and cast) or by leveraging the wisdom of the crowd (e.g., tag), and as such, they are prone to noise and are expensive to collect.  
Moreover, these features are often rare or absent for new items, making it difficult or even impossible to provide good quality recommendations. 

In this paper, we show that user's preferences on movies can be better described in terms of the \emph{\Mise}  features, i.e., the visual aspects of a movie that characterize design, aesthetics and style  (e.g., colors, textures). 
We use both {\it MPEG-7} visual descriptors and {\it Deep Learning} hidden layers as example of  \mise features that can visually describe movies. 
Interestingly, \mise features can be computed automatically from video files or even from trailers, offering more flexibility in handling new items, avoiding the need for costly and error-prone human-based tagging, and providing good scalability. 

We have conducted a set of experiments on a large catalogue of   4K movies.
Results show that recommendations based on \mise features consistently provide the best performance with respect to richer sets of more traditional features, such as genre and tag. 
 





\end{abstract}

%% file: acm-generated-code.tex
%
%

\ccsdesc[500]{Computer systems organization~Embedded systems}
\ccsdesc[300]{Computer systems organization~Redundancy}
\ccsdesc{Computer systems organization~Robotics}
\ccsdesc[100]{Networks~Network reliability}

%
%

%% file: introduction.tex
\section{Introduction}

Multimedia recommender systems base their recommendations on human-generated content features which are either crowd-sourced (\textit{e.g.},  tag) or editorial-generated (\textit{e.g.}, genre, director, cast).
The typical approach is to recommend items sharing features with the other items the user liked in the past.


In the movie domain, information about movies (\textit{e.g.}, tag, genre, cast) can be exploited to either use Content-Based Filtering (CBF) or to boost Collaborative Filtering (CF) with rich \emph{side information} \cite{shi2014collaborative}. A necessary prerequisite for both CBF and CF with side information is the availability of a rich set of descriptive \emph{features} about movies.

An open problem with multimedia recommender systems is how to enable or improve recommendations when user ratings and ``traditional'' human-generated features are nonexistent or incomplete. 
This is called the \emph{new item} problem  \cite{elahi2016survey,rubens2015active} and it happens frequently in video-on-demand scenarios, when new multimedia content is added to the catalog of available items   
(as an example, 500 hours of movie are uploaded to YouTube every minute\footnote{\url{http://www.reelseo.com/hours-minute-uploaded-youtube/}}).

Movie content features can be classified into three hierarchical levels \cite{wang2006video}.
\begin{itemize}

	\item  
At the highest level, we have \emph{semantic features} that deal with the conceptual model of a movie. 
An example of semantic feature is the plot of the movie \emph{The Good, the Bad and the Ugly}, which revolves around three gunslingers competing to find a buried cache of gold during the American Civil War;

	\item
At the intermediate level, we have \emph{syntactic features} that deal with objects in a movie and their interactions. 
As an example, in the same noted movie, there are Clint Eastwood, Lee Van Cleef, Eli Wallach, plus several horses and guns;

	\item
At the lowest level, we have \emph{stylistic features}, related to the \emph{\Mise} of the movie,
i.e., the design aspects that characterize aesthetic and style of a movie (e.g., colors or textures);
As an example, in the same movie predominant colors are yellow and brown, and camera shots use extreme close-up on actors' eyes. 

\end{itemize}
The same plot (semantic level) can be acted by different actors (syntactic level) and directed in different ways (stylistic level). 
In general, there is no direct link between the high-level concepts and the low-level features.
Each combination of features convey different communication effects and stimulate different feelings in the viewers.

Recommender systems in the movie domain mainly focus on high-level or intermediate-level features -- usually provided by a group of domain experts or by a large community of users -- such as movie genres (semantic features, high level), actors (syntactic features, intermediate level) or tags (semantic and syntactic features, high and intermediate levels) \cite{szomszor2007folksonomies,fleischman2003recommendations,jakob2009beyond}. 
Movie genres and actors are normally assigned by movie experts and tags by communities of users \cite{vig2009tagsplanations}. 
Human-generated features present a number of disadvantages: 
\begin{enumerate}
	\item features are prone to user biases and errors, therefore not fully reflecting the characteristics of a movie; 
	\item new items might lack features as well as ratings;
	\item unstructured features such as tags require complex Natural Language Processing (NLP) in order to account for stemming, stop words removal, synonyms detection and other semantic analysis tasks;
	\item not all features of an item have the same importance related to the task at hand; for instance, a background actor does not have the same importance as a guest star in defining the characteristics of a movie.
\end{enumerate} 

In contrast to human-generated features, the content of movie streams is itself a rich source of information about low-level stylistic features that can be used to provide movie recommendations. 
Low-level visual features have been shown to be very representative of the users feelings, according to the theory of \emph{Applied Media Aesthetics} \cite{ref:ZettlBook}.
By analyzing a movie stream content and extracting a set of low-level features, a recommender system can make personalized recommendations, tailored to a user's taste. 
This is particularly beneficial in the new item scenario, i.e., when movies without ratings and without user-generated tags are added to the catalogue. 

Moreover, while low-level visual features can be extracted from full-length movies, they can also be extracted from shorter version of the movies (i.e., trailers) in order to have a scalable recommender system.
I previous works, we have shown that \mise visual features extracted from trailers can be used to accurately predict genre of movies 
\cite{deldjoo2016content,deldjoo2015ecweb}.

In this paper, we show how to use low-level visual features extracted automatically from movie files as input to a hybrid CF+CBF algorithm.
We have extracted the low-level visual features by using two different approaches: 
\begin{itemize}
	\item \mpeg visual descriptors \cite{manjunath2001color}
	\item Pre-trained deep-learning neural networks (DNN) \cite{szegedy2015going}
\end{itemize}

Based on the discussion above, we articulate the following  research hypothesis: 
``\emph{a recommender system using low-level visual features (\mise) provides better accuracy compared to the same recommender system using traditional content features (genre and tag)}''.

We articulate the research hypothesis along the following  research questions: 
\begin{description}[noitemsep]
	\item[RQ1:] do visual low-level features extracted from any of \mpeg descriptors or pre-trained \deep provide better top-N recommendations than genre  and tag features?
	\item[RQ2:] do visual low-level features extracted from \mpeg descriptor in conjunction with pre-trained \deep provide better top-N recommendations than genre and tag features?
\end{description}

We have performed an exhaustive evaluation by comparing low-level visual features with respect to more traditional features (\textit{i.e.}, genre and tag). 
For each set of features, we have used a hybrid CBF+CF algorithm that includes item features as side information, where item similarity is learned with a Sparse LInear Method (SLIM) \cite{ning2012sparse}. 

We have used visual and content features either individually or in combination, in order to obtain a clear picture of the real ability of visual features in learning the preferences of users and effectively generating relevant recommendations.

We have computed different relevance metrics (precision, recall, and mean average precision) over a dataset of more than 8M ratings provided by 242K users to 4K movies. 
In our experiments, recommendations based on \mise visual features consistently provide the best performance.

Overall, this work provides a number of contributions to the RSs field in the movie domain: 
 
\begin{itemize}
	\item 
we propose a novel RS that automatically analyzes the content of the movies and extracts visual features in order to generate personalize recommendations for users;
	\item 
we evaluate recommendations by using a dataset of 4K movies and compare the results with a state-of-the-art hybrid CF+CBF algorithm;
	\item we have extracted \mise visual features adopting two different approaches (i.e., MPEG-7 and DNN) and fed them to the recommendation algorithm, either individually or in combination, in order to better study the power of these types of features; 
	\item 
the dataset, together with the user ratings and the visual features extracted from the videos (both \mpeg and deep-networks features), is available for download 
\footnote{ \url{recsys.deib.polimi.it}}.
\end{itemize}


The rest of the paper is organized as follows. 
Section~\ref{sec:related} reviews the relevant state of the art, related to content-based recommender  systems and video recommender systems. 
This section also introduces some theoretical background on Media Aesthetics that helps us to motivate our approach and interpret the results of our study. 
It describes the possible relation between the visual features adopted in our work and the aesthetic variables that are well known for artists in the domain of movie making.
In Section~\ref{sec:method_desc} we describe  our method for extracting and representing  \mise visual features of the movies and provide the details of our recommendation algorithms. 
Section \ref{sec:results}  introduces the evaluation method and presents the results of the study and Section~\ref{sec:discussion}  discusses them. Section~\ref{sec:conclusion}
draws the conclusions and identifies open issues and directions for future work.

%% file: related.tex
\section{Related work}
\label{sec:related}

\subsection{Multimedia Recommender Systems}

Multimedia recommender systems typically exploit {\it high-level} or {\it intermediate-level} features in order to generate movie recommendation~\cite{cantador2008enriching,musto2012enhanced}. 
This type of features express semantic and syntactic properties of media content that are obtained from structured sources of meta-information such as databases, lexicons and ontologies, or from less structured data such as reviews, news articles, item descriptions and social tags. 

In contrast, in this paper, we propose exploiting  {\it low-level} features to provide recommendations.
Such features express stylistic properties of the media content and are extracted directly from the multimedia content files \cite{deldjoo2016content}. 

While this approach has been already investigated in the music recommendation domain \cite{bogdanov2011unifying}, it has received marginal attention for movie recommendations. 
The very few approaches only consider low-level features to improve the quality of recommendations based on other type of features.
The work in~\cite{yang2007online} proposes a video  recommender system, called {\it VideoReach}, which incorporate  a combination of high-level and low-level video features (such as textual, visual and aural) in order to improve the click-through-rate metric.  
The work in~\cite{zhao2011integrating} proposes a multi-task learning algorithm to integrate multiple ranking lists, generated by using different sources of data, including visual content. 

While low-level features have been marginally explored in the community of recommender systems, they have been studied in other fields such as computer vision and content-based video retrieval.
The works in~\cite{hu2011survey,brezeale2008automatic} 
discuss a large body of low-level features (visual, auditory or textual) that can be considered for video content analysis. 
The work in~\cite{rasheed2005use} proposes a practical movie genre classification scheme based on computable visual cues.
~\cite{rasheed2003video} discusses a similar approach by considering also the audio features. 
Finally, the work in~\cite{zhou2010movie} proposes a framework for automatic classification of videos using visual features, based on the intermediate level of scene representation.

We note that, while the scenario of using the low-level features, as an additional side information, to hybridize the existing recommender systems is interesting, however, this paper addresses a different scenario, i.e., when the only available information is the low-level visual features
and the recommender system has to use it effectively for recommendation generation. 
Indeed, this is an extreme case of new item problem \cite{rubens2015active}, where traditional recommender systems fail in properly doing their job. It is worthwhile to note that while the present work has a focus on exploiting computer vision techniques on item description of products (i.e. the \textit{item-centric} aspect), computer vision techniques are also exploited in studying users' interaction behavior for example through studying their eye, gaze and head movement while navigating with a recommender system (i.e. the \textit{user-centric} aspect) \cite{bao2013your,chen2010eye,deldjoo2016low}.
\input{aesthetic}

%% file: aesthetic.tex
\subsection{Aesthetic View} 
\label{sec:artistic}

\begin{figure*}[ht!]
    \begin{center}
     \begin{tabular}{cc}  
				 \includegraphics[scale=0.66]{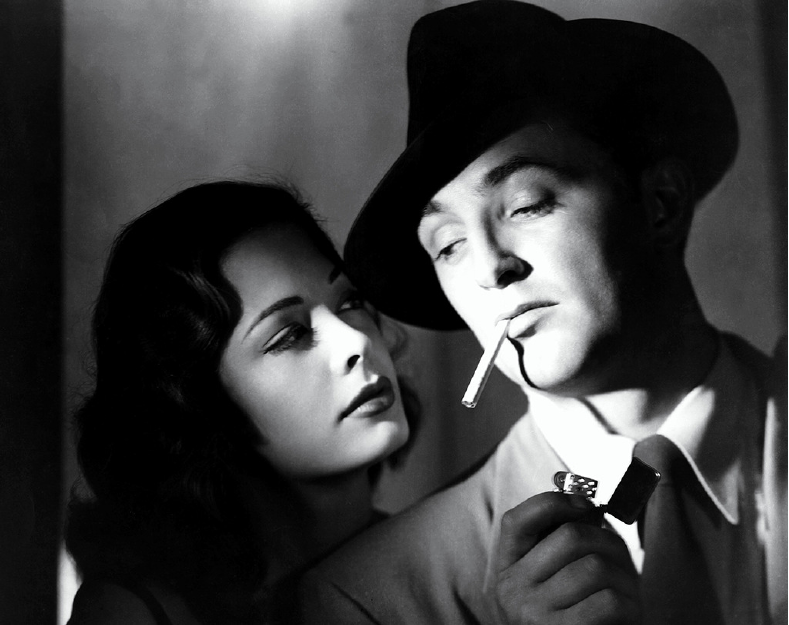}
 &
				 \includegraphics[scale=0.15]{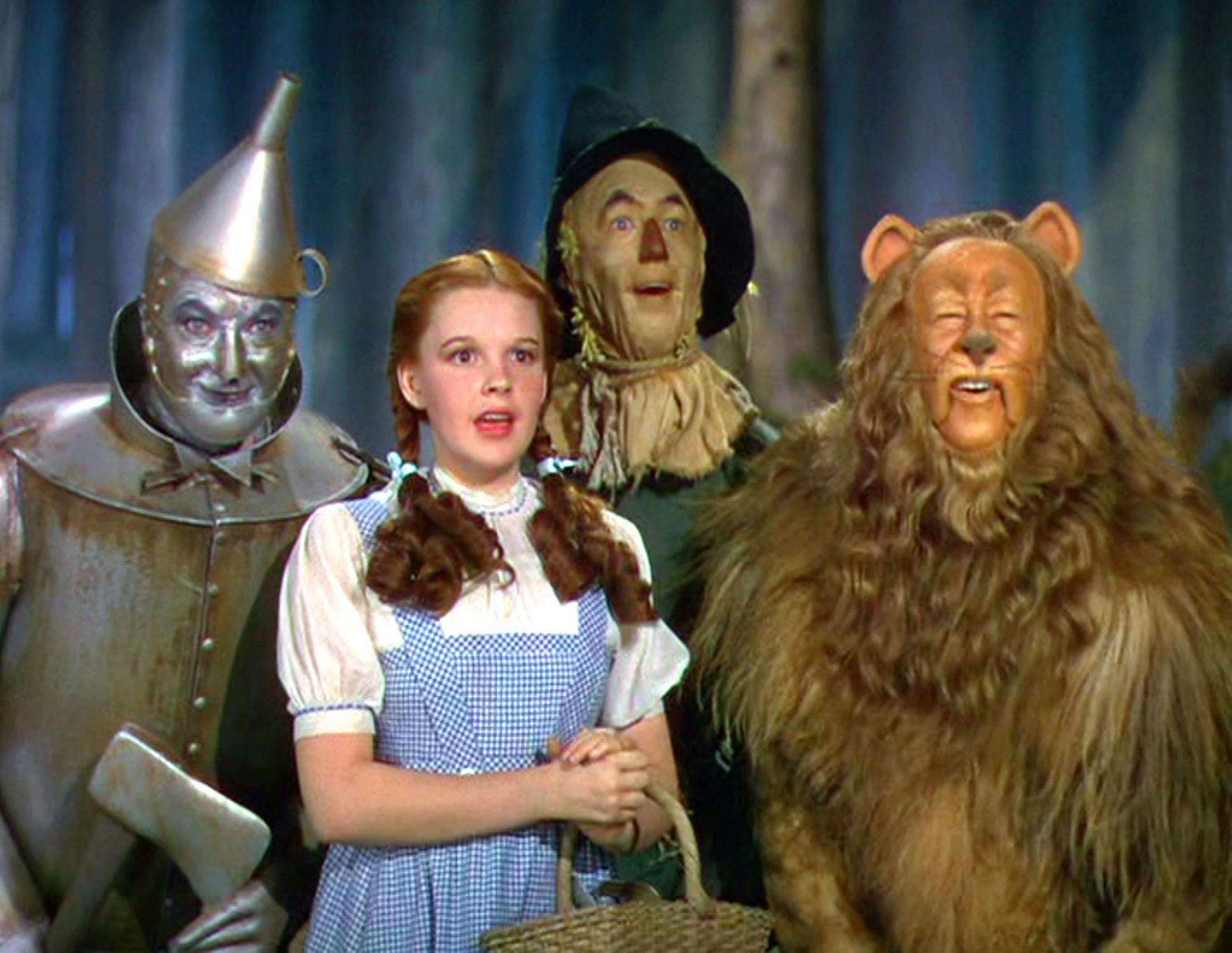} \\
		(a) & (b)\\
		\end{tabular}
		\caption{\textbf{a.} \textit{Out of the past} (1947) an example of highly contrasted lighting. \textbf{b.} \textit{The wizard of OZ} (1939) flat lighting example.}
		\label{figure:lighting}
	\end{center}
\end{figure*}

 \begin{figure*}[ht!]
    \begin{center}
     \begin{tabular}{cc}  
				 \includegraphics[scale=0.31]{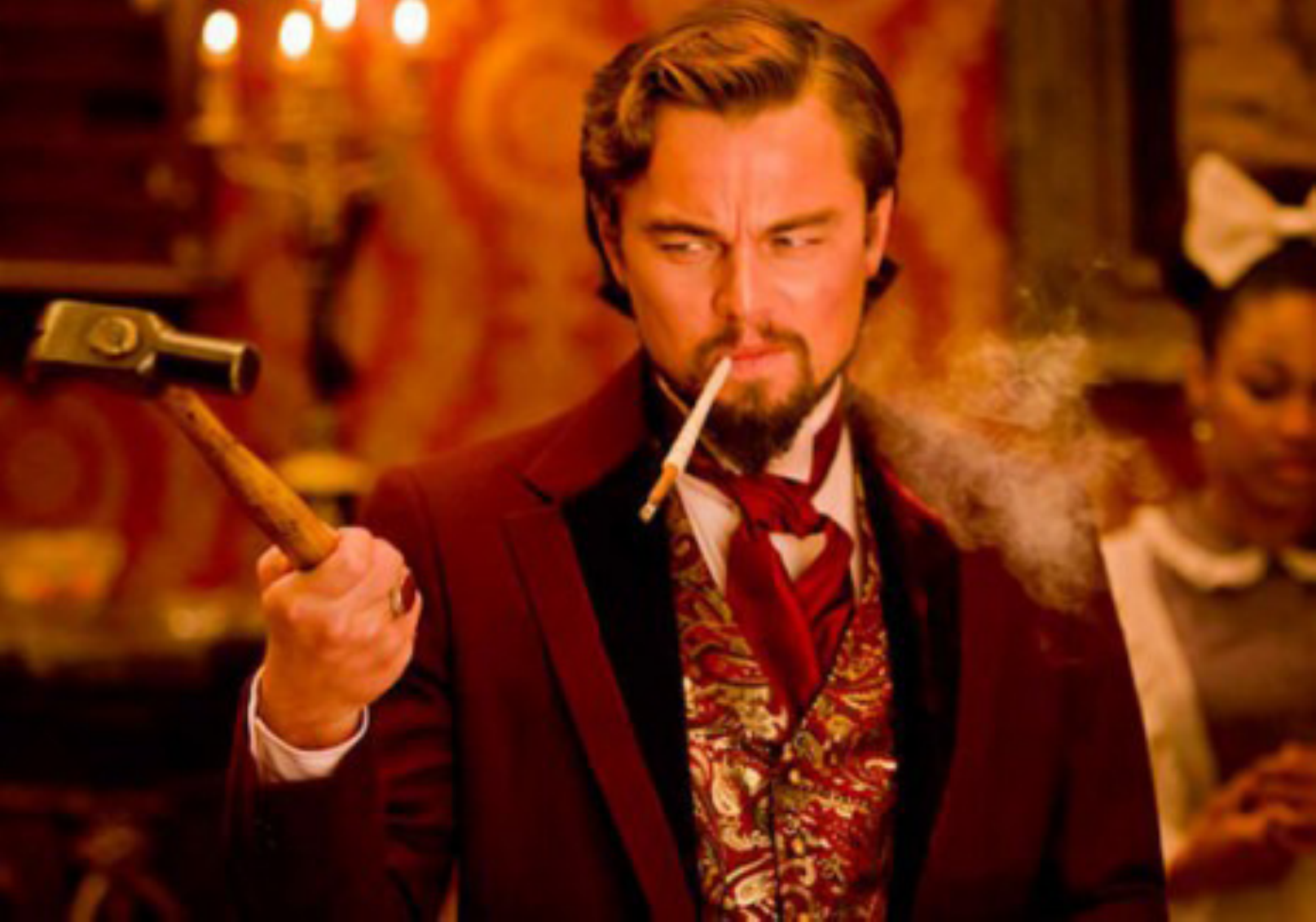}
 &
				 \includegraphics[scale=0.24]{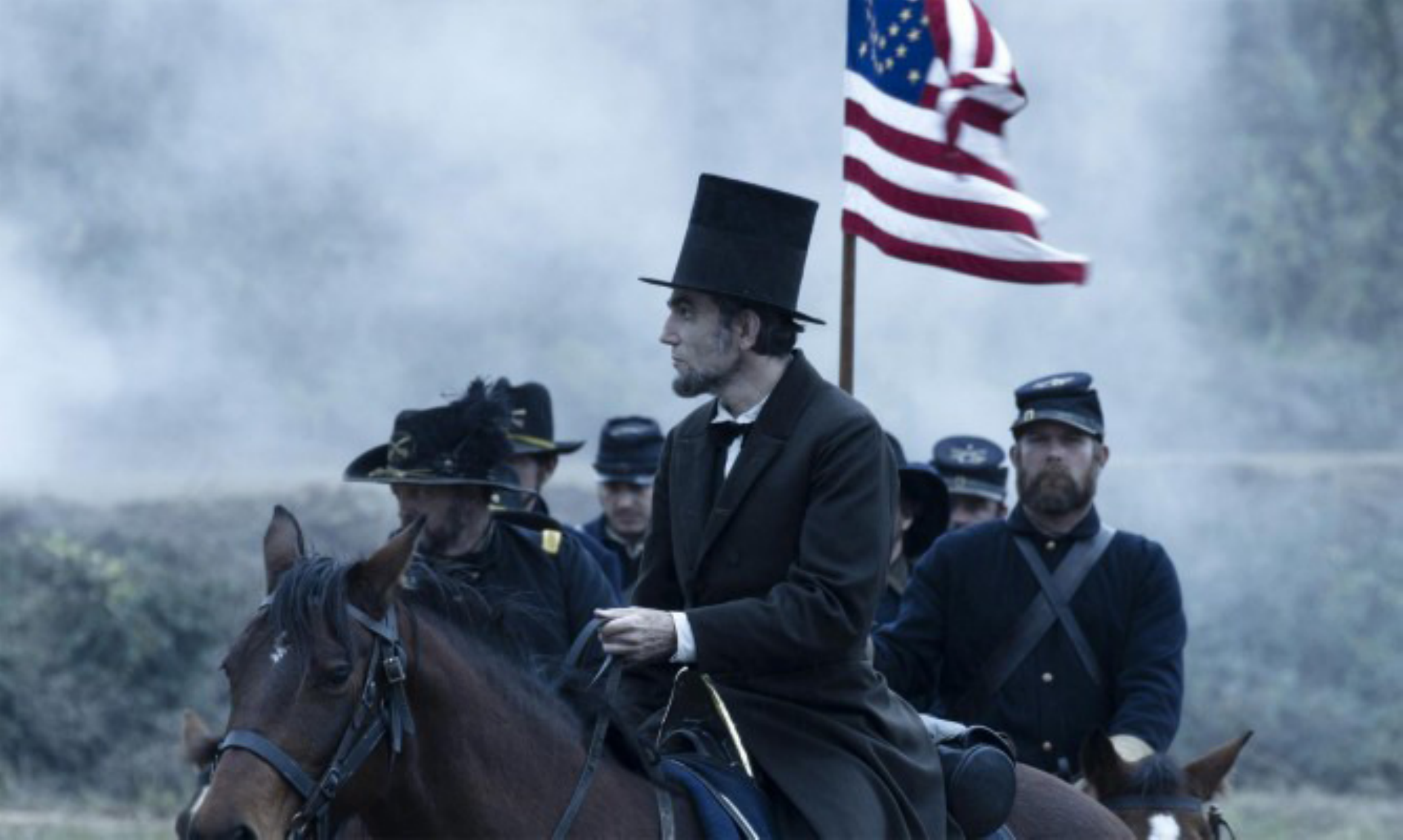}		\\
		(a) & (b)\\
		\end{tabular}
		\caption{\textbf{a.}  An image from \textit{Django Unchained} (2012). The red hue is used to increase the scene sense of violence. \textbf{b.} An image from \textit{Lincoln} (2012). Blue tone is used to produce the sense of coldness and fatigue experienced by the characters.}
		\label{figure:colors}
	\end{center}
\end{figure*}

The relation of \mise elements with the reactions they are able to evoke in viewers,
is one of the main concern of Applied Media Aesthetic \cite{ref:ZettlBook}.
Examples of \mise elements that are usually addressed in the literature on movie design are \emph{Lighting} and \emph{Color}   \cite{Dorai:2001}.

\emph{Lighting} is the deliberate manipulation of light for a certain communication purpose and it is used to create viewers' perception of the environment, and establish an aesthetic context for their experiences.
The two main lighting alternatives are usually addressed to as \emph{chiaroscuro} and \emph{flat lighting} \cite{zettl2013sight}. 
Figure \ref{figure:lighting}.a and Figure \ref{figure:lighting}.b exemplifies these two alternatives.

\emph{Colors} can strongly affect our perceptions and emotions in unsuspected ways.
For instance, red light gives the feeling of warmth, but also the feeling that time moves slowly,
while blue light gives the feeling of cold, but also that time moves faster.
The expressive quality of colors strongly depends on the lighting, since colors are a property of light  \cite{zettl2013sight}.
Figure \ref{figure:colors}.a and Figure \ref{figure:colors}.b present two examples of using colors in movies to evoke certain emotions.

Interestingly, most \mise elements can be computed from the video data stream as statistical values \cite{rasheed2005use,Buckland2008}. 
We call these computable aspects as {\it visual low level features} \cite{deldjoo2016recommending}.

%% file: method.tex
\section{Methodology}
\label{sec:method_desc}

The methodology adopted to provide recommendations based on visual features comprises \emph{five steps}:

\begin{enumerate}[noitemsep]

	\item 
\textbf{Video Segmentation}: the  goal is to segment each video into \emph{shots} and to select a representative key-frame from each shot;

	\item 
\textbf{Feature Extraction}: the  goal is to extract visual feature vectors from each key-frame. 
We have considered two different types of visual features for this purpose:
\begin{inlinelist}
  \item vectors extracted from \mpeg visual descriptors, and
  \item vectors extracted from pre-trained \deep ;
\end{inlinelist}

	\item 
\textbf{Feature Aggregation}: feature vectors extracted from the key-frame of a video are aggregated to obtain a feature vector descriptive of the whole video.
	\item 
\textbf{Feature Fusion}: in this step, features extracted from the same video but with different methods (e.g., \mpeg descriptors and \deep) are combined into a fixed-length descriptor;


	\item 
\textbf{Recommendation}: 
the (eventually aggregated) vectors describing low-level visual features of videos are used to feed a recommender algorithm.
For this purpose, we have considered the method \emph{Collective SLIM} as a feature-enhanced collaborative filtering (CF).

\end{enumerate}

\begin{figure*}
\begin{center}
\includegraphics[scale=0.45]{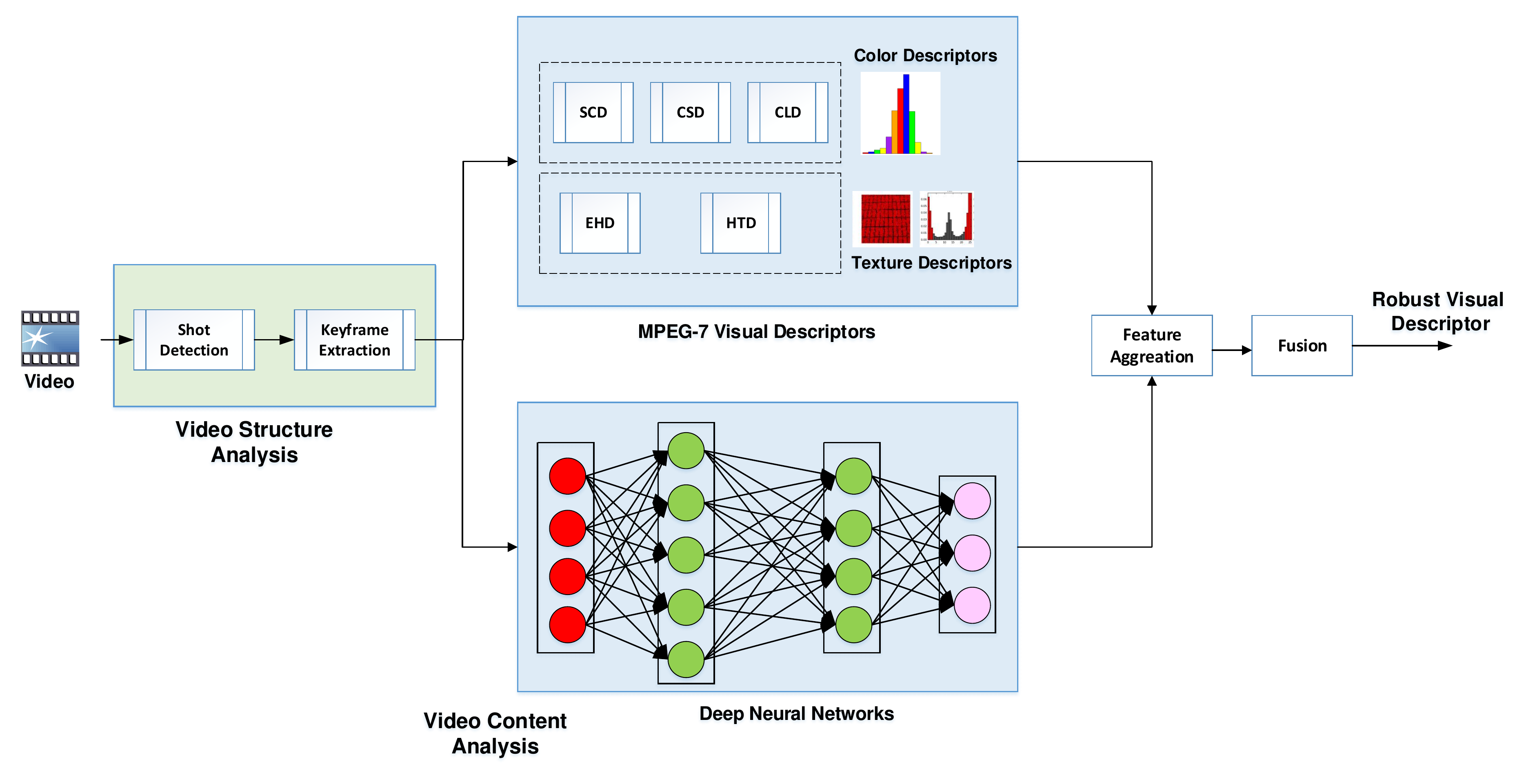}
\caption{Flowchart of the methdology used to create visual features from \mpeg descriptors and pre-trained \deep .}
\label{figure:flowchart}
\end{center}
\end{figure*}

The flowchart of the methodology is shown in Figure~\ref{figure:flowchart} and the 
steps are elaborated in more details in the following subsections.

\subsection{Video segmentation}
Shots are sequences of consecutive frames captured without interruption by a single camera. The transition between two successive shots of the video can be abrupt, where one frame belongs to a shot and the following frame belongs to the next shot, or gradual, where, two shots are combined using  chromatic, spatial or spatial-chromatic video production effects (e.g., fade in/out, dissolve, wipe), which gradually replace one shot by another. 

The color histogram distance is one of the most standard descriptors used as a measure of (dis)similarity between consecutive video frames in applications including: content-based video retrieval, object recognition, and others. A histogram is computed for each frame in the video and the~\emph{histogram intersection} is used as the means of comparing the local activity according to Equation~\ref{eq:shot_boundary},

\begin{equation}
\label{eq:shot_boundary}
\mathbf{s}(h_t,h_{t+1}) = \sum_{b}\min (h_t(b),h_{t+1}(b))
\end{equation}

where $h_t$ and $h_{t+1}$ are histograms of successive frames and $b$ is the index of the histogram bin. By comparing $\mathbf{s}$ with a predefined threshold, we segment the videos in our dataset into shots. We set the histogram similarity threshold to $0.75$.

\subsection{Feature extraction}
\label{sec:mpeg}

For each key frame, visual features are extracted by using either \mpeg descriptors or pre-trained \deep . 

\subsubsection{\mpeg features}

The \mpeg standard specifies descriptors that allow users to measure visual features of images.
More specifically, \mpeg specifies 17 descriptors divided into four categories: color, texture, shape, and motion \cite{manjunath2001color}.
In our work we have focused our attention on the following five \textbf{color} and \textbf{texture} descriptors, as previous experiments have proven the expressiveness of color and texture for similarity-based visual retrieval applications \cite{manjunath2001color,wang2014content}:
\begin{itemize}[noitemsep]
	\item \textbf{Color Descriptors.}
\begin{itemize}[noitemsep]
	\item
\emph{Scalable Color Descriptor (SCD)} is the color histogram of an image in the HSV color space. 
In our implementation we have used SCD with 256 coefficients (histogram bins). 
	\item
\emph{Color Structure Descriptor (CSD)} creates a modified version of the SCD histogram to take into account the physical position of each color inside the images, and thus it can capture both color content and information about the structure of this content. 
In our implementation, CSD is described by a feature vector of length 256.
	\item
\emph{Color Layout Descriptor (CLD)} is a very compact and resolution-invariant representation of color obtained by applying the DCT transformation on a 2-D array of representative colors in the YUV color space. 
CLD is described by a feature vector of length 120 in our implementation.
\end{itemize}
	\item \textbf{Texture Descriptors.}
\begin{itemize}[noitemsep]
	\item \emph{Edge Histogram Descriptor (EHD)} describes local \emph{edge distribution} in the frame. The image is divided into 16 non-overlapping blocks (subimages).
Edges within each block are classified into one of five edge categories: vertical, horizontal, left diagonal, right diagonal and non--directional edges. 
The final local edge descriptor is composed of a histogram with 5 x 16 = 80 histogram bins.
     \item \emph{Homogeneous Texture Descriptor (HTD)} describes homogeneous texture regions within a frame, by using a vector of 62 energy values. 
\end{itemize}
\end{itemize}

\subsubsection{\Deep features}

An alternative way to extract visual features from an image is to use the inner layers of pre-trained \deep \cite{he2015vbpr}.
We have used the 1024 inner neurons of GoogLeNet, a 22 layers deep network trained to classify over 1.2 million images classified into 1000 categories \cite{szegedy2015going}.
Each key frame is provided as input to the network and the activation values of inner neurons are used as visual features for the frame.

\subsection{Feature aggregation}

The previous step extracts a vector of features from each key-frame of a video.
We need to define a function to aggregate all these vectors into a single feature vector descriptive of the whole video.
The MPEG-7 standard defines an~\emph{extension} of the descriptors to a collection of pictures known as \emph{group of pictures} descriptors~\cite{manjunath2002introduction,bastan2010bilvideo}. 
The main aggregation functions are \emph{intersection histogram},
 \emph{average} and \emph{median}. 
Inspired by this, our proposed aggregation functions consist of the following:
\begin{itemize}[noitemsep]
	\item 
\textbf{intersection histogram}: each element of the aggregated feature vector is the~\emph{minimum} of the corresponding elements of the feature vectors from each key-frame;
\item 
\textbf{average}: each element of the aggregated feature vector is the~\emph{average} of the corresponding elements of the feature vectors from key-frame;
	\item
\textbf{median}: each element of the aggregated feature vector is the~\emph{median} of the corresponding elements of the feature vectors from key-frame.
	\item 
\textbf{union histogram}: each element of the aggregated feature vector is the~\emph{maximum} of the corresponding elements of the feature vectors from key-frame.
\end{itemize}

In our experiments we have applied each aggregation function to both \mpeg and deep-learning features.
\input{fusion}

\subsection{Recommendations}
\label{sec:algo}



\input{method-bpr}

%% file: fusion.tex
\subsection{Feature Fusion}
Motivated by the approach proposed in \cite{deldjoo2017combine}, we employed the fusion method based on~\emph{Canonical Correlation Analysis} (CCA) which exploits the low-level correlation between two set of visual features and learns a linear transformation that maximizes the pairwise correlation between two set of \mpeg and \deep visual features.

%% file: method-bpr.tex
In order to test the effectiveness of low-level visual features in video recommendations, we have experimented with a widely adopted  hybrid collaborative-filtering algorithm enriched with side information.

We use \emph{Collective SLIM} (Sparse Linear Method), a widely adopted sparse CF method that includes item features as side information to improve quality of recommendations \cite{ning2012sparse}. 
The item similarity matrix $S$ is learned by minimizing the following optimization problem
\begin{equation}
		\argmin_{S}{ \alpha \norm{R-RS} + \left(1 - \alpha \right) \norm{F-FS} + \gamma \norm{S} }
\end{equation}
where $R$ is the user-rating matrix, $F$ is the feature-item matrix, and parameters $\alpha$ and $\gamma$ are tuned with cross validation.
The algorithm is trained using Bayesian Pairwise Ranking \cite{rendle2009bpr}.

%% file: results.tex
\section{Evaluation and Results}
\label{sec:results}

\input{evaluation}

We feed the recommendation algorithm with \mpeg features, \deep features, genres and tags (tags are preprocessed with LSA).
We also feed the algorithm with a combinations of \mpeg and \deep features.

Tables \ref{table-3} present the results of the experiments in terms of precision, recall, and MAP, for different cutoff values, i.e., 1, 10, and 20.
First of all, we note that our initial analysis have shown that the best recommendation results are obtained by the intersection aggregation function for \mpeg and average for \deep features. Hence, we report the results for these two aggregation functions.

As it can be seen,  in terms of almost all the considered metrics, and all the cutoff values, \mpeg + DNN has shown the best results, and \mpeg alone has shown the second best results. The only exceptions are Precision at 10 and MAP at 10. In the former case, \mpeg is the best and Genre is the second. In the latter case, \mpeg is the best and \mpeg + DNN is the second. Unexpectedly, the recommendation based on tag is always the worst method. 

These results are very promising and overall present the power of recommendation based on \mpeg features, used individually or in combination with DNN features. Indeed, the results show that recommendation based on \mpeg features   always outperform genre and tag based recommendations, and the combination of \mpeg features with \deep significantly improves the quality of the hybrid CF+CBF algorithm and provides the best recommendation results overall.

 \input{bpr-table}

%% file: evaluation.tex

We have used the latest version of the 20M Movielens dataset \cite{harper2015movielens}. 
For each movie in the Movielens dataset, the title has been automatically queried in YouTube to search for the trailer. 

The final dataset contains \textbf{8'931'665 ratings} and \textbf{586'994 tags} provided by \textbf{242'209 users} to \textbf{3'964 movies} (sparsity 99.06\%) classified along \textbf{19 genres}:  action, adventure, animation, children's, comedy, crime, documentary,  drama, fantasy, film-noir, horror, musical, mystery, romance, sci-fi, thriller, war, western, and unknown.  

For each movie, the corresponding video trailer is available.
Low-level features have been automatically extracted from the trailers according to the methodology described in the previous section. 
The dataset, together with trailers and low-level features, is available for download \footnote{\url{recsys.deib.polimi.it}}.


In order to  evaluate the effectiveness of low-level visual features, we have used two baseline set of features: genre and tag. 
We have used Latent Semantic Analysis (LSA) to pre-process the tag-item matrix in order to better exploit the implicit structure in the association between tags and items. The technique includes decomposing the tag-item matrix into a set of orthogonal factors whose linear combination approximates the original matrix \cite{cremonesi2011looking}.


We evaluate the Top-$N$ recommendation quality by adopting a procedure similar to the one described in \cite{cremonesiKT10}.
\begin{itemize}[noitemsep]
	\item 
We randomly placed 80\% of the ratings in the training set, 10\% in the validation set, and
10\% in the test set.
Additionally, we performed a 5-fold cross validation test to compute confidence intervals.
\item For each relevant item $i$ rated by user $u$ in the test set, we form a list containing the item $i$ and all the items not rated by the user $u$, which we assume to be irrelevant to her. 
Then, we form a recommendation list by picking the top-$N$ ranked items. 
Being $r$ the rank of $i$, we have a $hit$ if $r<N$, otherwise we have a $miss$. 
	\item
We measure the quality of the recommendations in terms of \textbf{recall}, \textbf{precision} and \textbf{mean average precision (MAP)} for different cutoff values $N = 1, 10, 20$. 
\end{itemize}

%% file: bpr-table.tex
\begin{table*}
\centering
 \caption{Recommendation based on MPEG-7 and DNN features in comparison with the traditional genre  and tag features
 }
\label{table-3}

\begin{tabular}{ | l | c | c | c | c | c | c | c | c | c | }
\hline
	 & \multicolumn{3}{c|}{Recall}   & \multicolumn{3}{c|}{Precision}     & \multicolumn{3}{c|}{MAP}   \\ 
	\textbf{Features} & 1 & 10 & 20 & 1 & 10 & 20 & 1 & 10 & 20 \\ \hline
	MPEG-7 & 0.0337 & 0.1172 & 0.1751 & 0.1785 & \textbf{0.1354} & 0.1060 & 0.1785 & \textbf{0.1114} & 0.1001 \\ 
	DNN & 0.0238 & 0.0872 & 0.1381 & 0.1346 & 0.1021 & 0.0831 & 0.1346 & 0.0785 & 0.0714 \\ 
	MPEG-7 + DNN & \textbf{0.0383} & \textbf{0.1773} & \textbf{0.2466} & \textbf{0.2005} & 0.1063 & \textbf{0.0771} & \textbf{0.2005} & 0.1051 & \textbf{0.1040} \\ 
	Genre & 0.0294 & 0.0904 & 0.1381 & 0.1596 & 0.1108 & 0.0892 & 0.1596 & 0.0892 & 0.0792 \\ 
	Tag-LSA & 0.0127 & 0.0476 & 0.0684 & 0.1001 & 0.0686 & 0.0523 & 0.1001 & 0.0493 & 0.0389 \\ \hline
\end{tabular}
\end{table*}

%% file: discussion.tex
\section{Discussion}
\label{sec:discussion} 

Our results provide empirical evidence that visual low-level features extracted from \mpeg descriptors provide better top-N recommendations than genres and tags while the same does not apply to visual low-level features extracted from pre-trained deep-learning networks.

Overall, our experiments prove the effectiveness of movie recommendations based on  visual features automatically extracted from trailers of movies. 
Recommendations based on \deep visual features can provide good quality recommendations, in line with recommendations based on human-generated attributes such as genres and tags while visual features extracted from \mpeg descriptors consistently provide better recommendations. Moreover, fusion of the \deep and MPEG-7 visual features performs the best recommendation results. These results suggest an interesting consideration: \emph{users' opinions on movies are influenced more by style than content}.

Given the small number of \mise features (e.g., the combined \mpeg feature vector contains 774 elements only, compared with half a million tags) and the fact that we extracted them from movie trailers, we did not expect this result. 
In fact, we would view it as a good news for practitioners of movie recommender systems, as low-level features combine multiple advantages. 
First, \mise features have the convenience of being computed automatically from video files, offering designers more flexibility in handling new items, without the need to wait for costly editorial or crowd-based tagging.
Moreover, it is also possible to extract low level features from movie trailers, without the need to work on full-length videos \cite{deldjoo2016content}.
This guarantees good scalability. 
Finally, viewers are less consciously aware of movie styles and we expect that recommendations based on low level features could to be more attractive in terms of diversity, novelty and serendipity.

We would like to offer an explanation as to why \mise low-level features consistently deliver better top-N recommendations than a much large number of high-level attributes. 
This may have to do with a limitation of high-level features, which are binary in nature: movies either have or not have a specific attribute. 
On the contrary low-level features are continuous in their values and they are present in all movies, but with different weights.

A potential difficulty in exploiting \mise low-level visual features is the computational load required for the extraction of features from full-length movies.  However, we have observed that low-level visual features extracted from the movie trailers are highly correlated with the corresponding features extracted from full-length movies \cite{deldjoo2016content}. Accordingly, the observed strong correlation indicates that the movie trailers    
are indeed perfect representatives of the corresponding full-length movies. Hence, instead of analyzing the lengthy full 
movies the trailers can be properly analyzed which can result in significant reduction in the  computational load  of using \mise low-level 
visual features.

%% file: conclusion.tex
\section{Conclusion and Future Work}
\label{sec:conclusion} 

This work presents a novel approach in the domain of movie recommendations.
The technique is based on the analysis of movie content and extraction of stylistic low-level 
features that are used to generate personalized recommendations for users.
This approach makes it possible to recommend items to users without relying on any high-level semantic features (such as, genre, or tag) 
that are expensive to obtain, as they require expert level knowledge, and shall be missing (e.g., in new item scenario).


While the results of this study would not underestimate the importance of high-level semantic 
features, however, they provide a strong argument for exploring the potential of low-level
visual features that are automatically extracted from movie content. 

For future work, we consider the design and development of an online web application in order to conduct online studies with real user. 
The goal is to evaluate the effectiveness of recommendations based on low-level visual features not only in terms of relevance, but also in terms of novelty and diversity. 
Moreover, we will extend the range of low-level features extracted, and also, include audio features. 	
We will also extend the evaluation to user-generated videos (e.g., YouTube).
Finally, we would feed the MPEG-7 features as input to the initial layer of deep-networks and build the model accordingly.
We are indeed, interested to investigate the possible  improvement of recommendation based on the features
provided by the deep-networks trained in this way.
 